\documentclass[]{aastex63}
\usepackage{amssymb, amsmath,framed}
\usepackage{gensymb}
\usepackage{multirow}
\usepackage{rotating}
\usepackage{savesym}
\savesymbol{tablenum}
\usepackage{siunitx}
\restoresymbol{SIX}{tablenum}

\usepackage{afterpage}

\usepackage{bm}
\expandafter\ifx\csname package@font\endcsname\relax\else
 \expandafter\expandafter
 \expandafter\usepackage
 \expandafter\expandafter
 \expandafter{\csname package@font\endcsname}

\fi
\def\bq{\begin{equation}}
\def\eq{\end{equation}}
\def\bqy{\begin{eqnarray}}
\def\eqy{\end{eqnarray}}

\def\p{\partial}

\def\p{\partial}

\def\R{\mathbb{R}}

\usepackage{atbegshi}

\begin{document}
\title{\large{Analysis of Low $\Delta V$ Spacecraft Missions to Oort Cloud Comet C/2014 UN$_{271}$}}

\correspondingauthor{Adam Hibberd}
\email{adam.hibberd@i4is.org}

\author{Adam Hibberd}
\affiliation{Initiative for Interstellar Studies (i4is) 27/29 South Lambeth Road London, SW8 1SZ United Kingdom}

\author{T. Marshall Eubanks}
\affiliation{Space Initiatives Inc, Newport, VA 24128, USA}

\begin{abstract}
Comet C/2014 UN$_{271}$, alternative designation \emph{BB} after its discoverers \emph{Bernardinelli–Bernstein}, and commonly referred to as UN$_{271}$, is an extreme case on two fronts, firstly its solar distance on discovery ($>$ 29 $\si{au}$) and secondly the size of its nucleus (137±15 $\si{km}$). 
With an aphelion distance of $\sim$33,000 au (w.r.t. the solar system barycentre) and an orbital period $\sim$2 million years, it is definitely an object from the solar system's `Oort cloud', and also by a good measure the largest Oort cloud object ever observed. \emph{In situ} observation of UN$_{271}$ would be of considerable scientific importance. Unlike most Oort cloud comets which have been discovered for the first time only as they near the inner solar system,  UN$_{271}$ was discovered early enough to provide adequate advanced warning to plan for such a mission. In this paper we describe the various methods for reaching UN$_{271}$ during the period around its perihelion and ecliptic plane passage, with both flyby and rendezvous options; exploiting direct transfers, Jupiter powered gravitational assists (GA) or alternatively a series of GAs of the inner planets. Viable flyby and rendezvous trajectories are found, especially using the NASA Space Launch System (SLS) as the launch vehicle.   
\end{abstract}

\section{Introduction} \label{SecIntro}

C/2014 UN$_{271}$ \citep{Bernardinelli_2022}: by far the largest Oort cloud object known, with an effective diameter of 137±15 km \citep{refId0}, UN$_{271}$ will not come closer than the orbit of Saturn, with perihelion passage in 2031 at 11.0 au and ecliptic intersection on August 7, 2033, at a solar distance of 12.0 au. It is also unusual in that it exhibited activity well before perihelion, at a distance of 23.8 au away from the sun when the surface was still quite cold, well before the water frost line of $\sim$5 au \citep{Farnham_2021}. The orbital simulations of  \cite{Bernardinelli_2021} suggest there was at least one previous perihelion passage, the last approaching a distance of 18 au from the sun. The primary science question of any mission to UN$_{271}$ would be to determine where, and how, did this Oort cloud object form?\\
As far as a mission is concerned, an \emph{in situ} spacecraft exploration would be ideal for answering these questions, however this comet is way outside of the range of the forthcoming European Space Agency (ESA) \emph{Comet Interceptor} spacecraft \citep{S_nchez_2021} to be launched in 2029. Thus this paper examines the general feasibility of missions to UN$_{271}$ around the time frame of its perihelion passage, so in the '20s to '30s, with particular attention to chemical propulsion, so high thrust. `General feasibility' in this context is equivalent to `minimum $\Delta V$', this being the objective of the analysis conducted here and the precise definition of this parameter will be articulated below.\\

\section{Method} \label{method}

Normally for spacecraft mission preliminary design, we seek to minimize fuel usage and so maximize the useful payload mass. This general aim, which would in practice have to be studied in detail on a specific mission-by-mission basis, can nevertheless be addressed in a generic sense by various methods and adopting different assumptions.\\
For this study, high thrust propulsion methods such as chemical rockets are assumed. This allows one to approximate the spacecraft trajectory as a sequence of conic-sections connected by discrete instantaneous impulsive changes in velocity ($\Delta V$) at each of the nodes (therefore supposing infinite and instantaneous thrust). Furthermore if a node is a planet, then a gravitational assist (GA) may be used to accelerate or decelerate the craft with respect to the sun, with no requirement for propellant. 
The effectiveness of this GA may be further amplified by an application of $\Delta V$ at the periapsis of the spacecraft - its closest approach to the planet in question. (The periapsis point is chosen to receive maximum benefit from the Oberth effect \citep{doi:10.1119/10.0001956}.) \\
The periapsis and $\Delta V$ for each planetary encounter can be computed by assuming the arrival velocity of the preceding conic-section and the departure velocity of the subsequent conic-section (both  w.r.t. the planet) are the hyperbolic arrival and departure excess velocities respectively. This permits calculation of two hyperbolas connected at a common peripasis point, the corresponding $\Delta V$ being simply that delivered at periapsis to get from the first to the second. All this is modelled by the preliminary interplanetary mission design software known as \textit{Optimum Interplanetary Trajectory Software} (OITS), refer \cite{OITS_info}. This software uses two Non-Linear Programming (NLP) solvers, NOMAD \citep{LeDigabel2011} and MIDACO \citep{Schlueter_et_al_2013,Schlueter_et_al_2009}.\\
Further explanation of OITS and elucidation on a feature known as \textit{'Intermediate Points'}, which can be used to model $V_{\infty}$ Leveraging Manoeuvres \citep{doi:10.2514/6.1994-3769}, can be found at \cite{https://doi.org/10.48550/arxiv.2205.10220}.
Note the theory applied above restricts this study to general feasibility and trajectory design considerations as the assumptions ignore additional forces influencing the spacecraft's motion (such as solar radiation pressure). However this is not the case for the celestial bodies visited as for this analysis, extremely accurate ephemerides of the planets and for that matter UN$_{271}$ were furnished via the NASA JPL Horizons service, and appropriate NASA SPICE kernels were generated and utilised by OITS. For UN$_{271}$, the kernel file was generated in June 2021.

\section{Results} \label{results}
\subsection{Flyby Missions} \label{flyby_res}
Flyby missions to UN$_{271}$, i.e. missions which approach and depart the objective without any significant correction to match velocities with the target, may be achieved by direct transfer from Earth to UN$_{271}$ or using a combination of powered or unpowered GAs.
\subsubsection{Direct Transfer} \label{flyby_dir}
In this section we assume a direct transfer to UN$_{271}$ and that we wish to minimize the Hyperbolic Excess Speed at Earth, designated $V_{\infty}$. Generally, in preliminary mission design of the kind adopted here, the task is to minimize some metric of the interplanetary trajectory, usually $\Delta V$, the definition of this depending on the precise context. For the moment we shall assume $\Delta V$ $=$ $V_{\infty}$.
\begin{figure}[ht]
\hspace{0.7cm}
\includegraphics[scale=0.50]{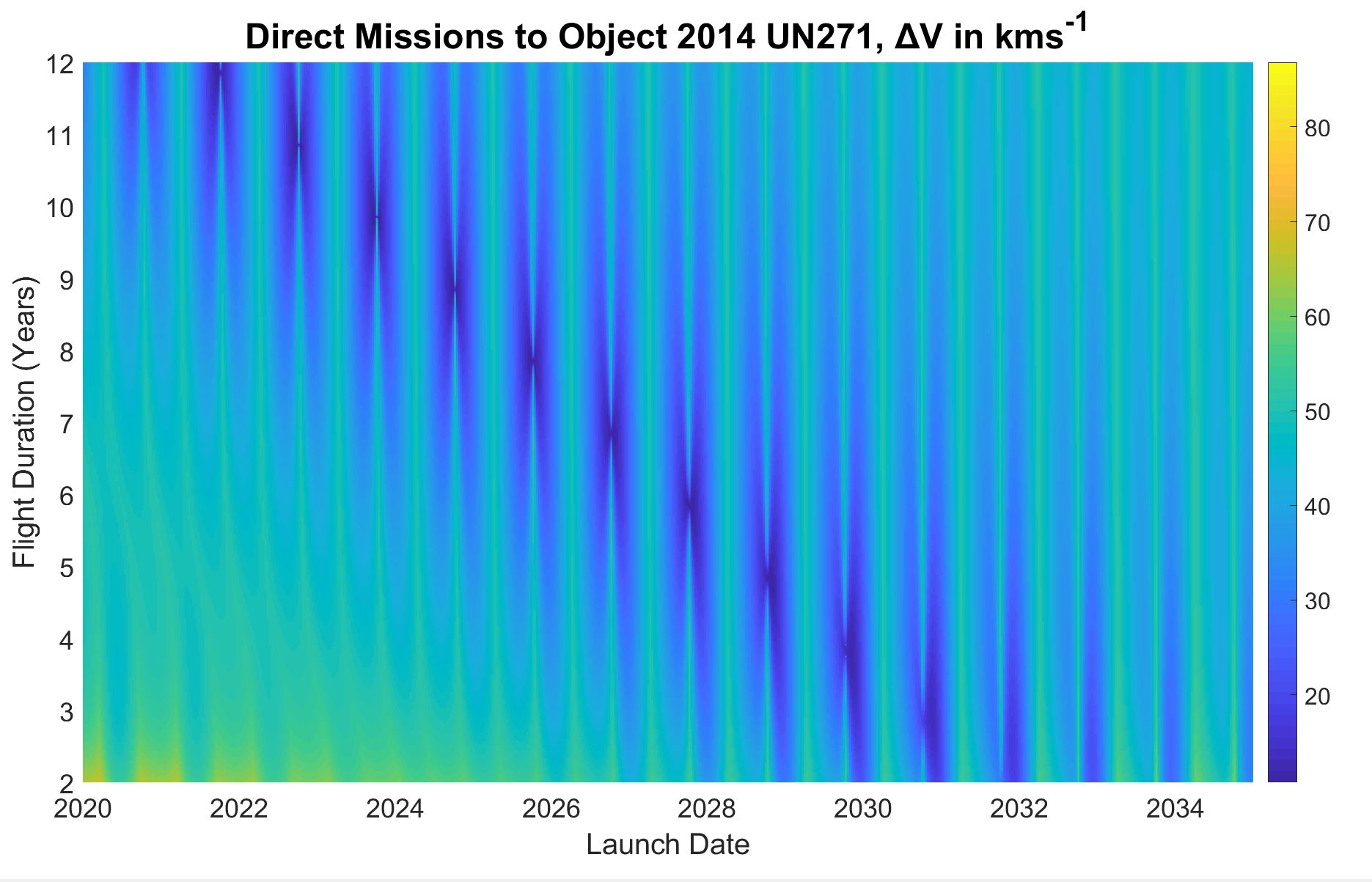}
\caption{Pork Chop Plot for Flyby Missions to UN$_{271}$ with colour bar to the right with units in $\si{km.s^{-1}}$.}
\label{fig:Direct}
\end{figure}
Figure \ref{fig:Direct} is a colour contour plot of direct trajectories, with two independent variables – on the x-axis are launch dates between 2020 and 2035 and on the y-axis are overall flight durations in years. The darker and bluer the colours, the lower the $\Delta V$. Such plots are often referred to as \emph{pork chop plots}. We see in this contour plot the yearly patterns resulting from the Earth occupying yearly sweet spots in its orbit with respect to UN$_{271}$, at which points the relative alignment of the two bodies are particularly propitious for missions.\\

\begin{table*}[ht]
\caption{Direct Missions to UN$_{271}$ with flyby at target}
\label{tab:Direct_T}
\begin{tabular}{llllllll}
\hline
\multicolumn{1}{|c|}{\textbf{Launch}} &
  \multicolumn{1}{c|}{\textbf{Arrival}} &
  \multicolumn{1}{c|}{\textbf{Flight}} &
  \multicolumn{1}{c|}{\textbf{Flight}} &
  \multicolumn{1}{c|}{\textbf{$\Delta V$ at}} &
  \multicolumn{1}{c|}{\textbf{$C_3$ at}} &
  \multicolumn{1}{c|}{\textbf{Arrival}} &
  \multicolumn{1}{c|}{\textbf{Phase}} \\
\multicolumn{1}{|c|}{\textbf{Date}} &
  \multicolumn{1}{c|}{\textbf{Date}} &
  \multicolumn{1}{c|}{\textbf{Duration}} &
  \multicolumn{1}{c|}{\textbf{Duration}} &
  \multicolumn{1}{c|}{\textbf{Earth}} &
  \multicolumn{1}{c|}{\textbf{Earth}} &
  \multicolumn{1}{c|}{\textbf{Velocity}} &
  \multicolumn{1}{c|}{\textbf{Angle}} \\
\multicolumn{1}{|c|}{} &
  \multicolumn{1}{c|}{} &
  \multicolumn{1}{c|}{\textbf{(Days)}} &
  \multicolumn{1}{c|}{\textbf{(Years)}} &
  \multicolumn{1}{c|}{\textbf{($\si{km.s^{-1}}$)}} &
  \multicolumn{1}{c|}{\textbf{($\si{km^2.s^{-2}}$)}} &
  \multicolumn{1}{c|}{\textbf{($\si{km.s^{-1}}$)}} &
  \multicolumn{1}{c|}{\textbf{(degs)}} \\ \hline
\multicolumn{1}{|c|}{24-Sep-22} &
  \multicolumn{1}{c|}{06-Aug-33} &
  \multicolumn{1}{c|}{3969} &
  \multicolumn{1}{c|}{10.87} &
  \multicolumn{1}{c|}{10.77} &
  \multicolumn{1}{c|}{115.92} &
  \multicolumn{1}{c|}{13.74} &
  \multicolumn{1}{c|}{114.89} \\ \hline
\multicolumn{1}{|c|}{27-Sep-23} &
  \multicolumn{1}{c|}{06-Aug-33} &
  \multicolumn{1}{c|}{3601} &
  \multicolumn{1}{c|}{9.86} &
  \multicolumn{1}{c|}{10.73} &
  \multicolumn{1}{c|}{115.07} &
  \multicolumn{1}{c|}{13.44} &
  \multicolumn{1}{c|}{112.05} \\ \hline
\multicolumn{1}{|c|}{29-Sep-24} &
  \multicolumn{1}{c|}{06-Aug-33} &
  \multicolumn{1}{c|}{3233} &
  \multicolumn{1}{c|}{8.85} &
  \multicolumn{1}{c|}{10.7} &
  \multicolumn{1}{c|}{114.43} &
  \multicolumn{1}{c|}{13.13} &
  \multicolumn{1}{c|}{108.47} \\ \hline
\multicolumn{1}{|c|}{02-Oct-25} &
  \multicolumn{1}{c|}{06-Aug-33} &
  \multicolumn{1}{c|}{2865} &
  \multicolumn{1}{c|}{7.84} &
  \multicolumn{1}{c|}{10.69} &
  \multicolumn{1}{c|}{114.22} &
  \multicolumn{1}{c|}{12.83} &
  \multicolumn{1}{c|}{103.83} \\ \hline
\multicolumn{1}{|c|}{07-Oct-26} &
  \multicolumn{1}{c|}{06-Aug-33} &
  \multicolumn{1}{c|}{2495} &
  \multicolumn{1}{c|}{6.83} &
  \multicolumn{1}{c|}{10.72} &
  \multicolumn{1}{c|}{114.87} &
  \multicolumn{1}{c|}{12.57} &
  \multicolumn{1}{c|}{97.64} \\ \hline
\multicolumn{1}{|c|}{13-Oct-27} &
  \multicolumn{1}{c|}{06-Aug-33} &
  \multicolumn{1}{c|}{2124} &
  \multicolumn{1}{c|}{5.82} &
  \multicolumn{1}{c|}{10.83} &
  \multicolumn{1}{c|}{117.32} &
  \multicolumn{1}{c|}{12.46} &
  \multicolumn{1}{c|}{89.21} \\ \hline
\multicolumn{1}{|c|}{20-Oct-28} &
  \multicolumn{1}{c|}{07-Aug-33} &
  \multicolumn{1}{c|}{1752} &
  \multicolumn{1}{c|}{4.8} &
  \multicolumn{1}{c|}{11.13} &
  \multicolumn{1}{c|}{123.85} &
  \multicolumn{1}{c|}{12.77} &
  \multicolumn{1}{c|}{77.66} \\ \hline
\multicolumn{1}{|c|}{31-Oct-29} &
  \multicolumn{1}{c|}{19-Aug-33} &
  \multicolumn{1}{c|}{1388} &
  \multicolumn{1}{c|}{3.8} &
  \multicolumn{1}{c|}{11.86} &
  \multicolumn{1}{c|}{140.72} &
  \multicolumn{1}{c|}{14.07} &
  \multicolumn{1}{c|}{62.72} \\ \hline
\multicolumn{1}{|c|}{13-Nov-30} &
  \multicolumn{1}{c|}{25-Oct-33} &
  \multicolumn{1}{c|}{1077} &
  \multicolumn{1}{c|}{2.95} &
  \multicolumn{1}{c|}{13.63} &
  \multicolumn{1}{c|}{185.82} &
  \multicolumn{1}{c|}{17.02} &
  \multicolumn{1}{c|}{47.38} \\ \hline
\multicolumn{1}{|c|}{26-Nov-31} &
  \multicolumn{1}{c|}{12-May-34} &
  \multicolumn{1}{c|}{898} &
  \multicolumn{1}{c|}{2.46} &
  \multicolumn{1}{c|}{17.06} &
  \multicolumn{1}{c|}{290.97} &
  \multicolumn{1}{c|}{20.46} &
  \multicolumn{1}{c|}{37.01} \\ \hline

\end{tabular}
\end{table*}

\afterpage{\clearpage
\begin{sidewaystable*}[ht]
\centering
\caption{Indirect Missions to UN$_{271}$ with flyby at target}
\label{tab:InDirect_T}
\begin{tabular}{cccccccccc}
\hline
\multicolumn{1}{|c|}{} &
  \multicolumn{1}{c|}{Aphelia for} &
  \multicolumn{1}{c|}{} &
  \multicolumn{1}{c|}{} &
  \multicolumn{1}{c|}{\textbf{}} &
  \multicolumn{1}{c|}{} &
  \multicolumn{1}{c|}{} &
  \multicolumn{1}{c|}{} &
  \multicolumn{1}{c|}{} &
  \multicolumn{1}{c|}{} \\
\multicolumn{1}{|c|}{Trajectory} &
  \multicolumn{1}{c|}{Resonances} &
  \multicolumn{1}{c|}{} &
  \multicolumn{1}{c|}{} &
  \multicolumn{1}{c|}{\textbf{}} &
  \multicolumn{1}{c|}{} &
  \multicolumn{1}{c|}{Approach} &
  \multicolumn{1}{c|}{Arrival} &
  \multicolumn{1}{c|}{} &
  \multicolumn{1}{c|}{Phase} \\
\multicolumn{1}{|c|}{(r=resonance)} &
  \multicolumn{1}{c|}{and} &
  \multicolumn{1}{c|}{Launch} &
  \multicolumn{1}{c|}{Arrival} &
  \multicolumn{1}{c|}{\textbf{Total $\Delta V$}} &
  \multicolumn{1}{c|}{$C_3$ (\si{km^2.s^{-2}})} &
  \multicolumn{1}{c|}{Velocity} &
  \multicolumn{1}{c|}{Heliocentric} &
  \multicolumn{1}{c|}{In-flight $\Delta V$} &
  \multicolumn{1}{c|}{Angle} \\
\multicolumn{1}{|c|}{} &
  \multicolumn{1}{c|}{heliocentric} &
  \multicolumn{1}{c|}{Date} &
  \multicolumn{1}{c|}{Date} &
  \multicolumn{1}{c|}{\textbf{(\si{km.s^{-1}})}} &
  \multicolumn{1}{c|}{} &
  \multicolumn{1}{c|}{(\si{km.s^{-1}})} &
  \multicolumn{1}{c|}{Distance (au)} &
  \multicolumn{1}{c|}{(\si{km.s^{-1}})} &
  \multicolumn{1}{c|}{(deg)} \\
\multicolumn{1}{|c|}{} &
  \multicolumn{1}{c|}{distance of} &
  \multicolumn{1}{c|}{} &
  \multicolumn{1}{c|}{} &
  \multicolumn{1}{c|}{\textbf{}} &
  \multicolumn{1}{c|}{} &
  \multicolumn{1}{c|}{} &
  \multicolumn{1}{c|}{} &
  \multicolumn{1}{c|}{} &
  \multicolumn{1}{c|}{} \\
\multicolumn{1}{|c|}{} &
  \multicolumn{1}{c|}{DSM (au)} &
  \multicolumn{1}{c|}{} &
  \multicolumn{1}{c|}{} &
  \multicolumn{1}{c|}{\textbf{}} &
  \multicolumn{1}{c|}{} &
  \multicolumn{1}{c|}{} &
  \multicolumn{1}{c|}{} &
  \multicolumn{1}{c|}{} &
  \multicolumn{1}{c|}{} \\ \hline
\multicolumn{1}{|c|}{E-V-r-V-E-UN$_{271}$} &
  \multicolumn{1}{c|}{1.57} &
  \multicolumn{1}{c|}{2028 MAR 10} &
  \multicolumn{1}{c|}{2033 OCT 18} &
  \multicolumn{1}{c|}{\textbf{5.53}} &
  \multicolumn{1}{c|}{10.35} &
  \multicolumn{1}{c|}{13.5} &
  \multicolumn{1}{c|}{12.11} &
  \multicolumn{1}{c|}{2.32} &
  \multicolumn{1}{c|}{114.89} \\ \hline
\multicolumn{1}{|c|}{E-V-E-r-E-UN$_{271}$} &
  \multicolumn{1}{c|}{2.2} &
  \multicolumn{1}{c|}{2026 AUG 06} &
  \multicolumn{1}{c|}{2033 SEP 09} &
  \multicolumn{1}{c|}{\textbf{6.42}} &
  \multicolumn{1}{c|}{12.6} &
  \multicolumn{1}{c|}{14.01} &
  \multicolumn{1}{c|}{12.02} &
  \multicolumn{1}{c|}{2.87} &
  \multicolumn{1}{c|}{116.01} \\ \hline
\multicolumn{1}{|c|}{E-V-r-V-r-V-UN$_{271}$} &
  \multicolumn{1}{c|}{1.57, 2.92} &
  \multicolumn{1}{c|}{2025 JAN 07} &
  \multicolumn{1}{c|}{2033 NOV 12} &
  \multicolumn{1}{c|}{\textbf{6.5}} &
  \multicolumn{1}{c|}{13.7} &
  \multicolumn{1}{c|}{12.5} &
  \multicolumn{1}{c|}{12.16} &
  \multicolumn{1}{c|}{2.8} &
  \multicolumn{1}{c|}{103.9} \\ \hline
\multicolumn{1}{|c|}{E-V-r-V-r-V-UN$_{271}$} &
  \multicolumn{1}{c|}{1.57, 2.28} &
  \multicolumn{1}{c|}{2025 JAN 11} &
  \multicolumn{1}{c|}{2033 NOV 11} &
  \multicolumn{1}{c|}{\textbf{7.04}} &
  \multicolumn{1}{c|}{13.63} &
  \multicolumn{1}{c|}{12.2} &
  \multicolumn{1}{c|}{12.16} &
  \multicolumn{1}{c|}{3.34} &
  \multicolumn{1}{c|}{95.68} \\ \hline
\multicolumn{1}{|c|}{E-V-r-V-M-V-UN$_{271}$} &
  \multicolumn{1}{c|}{1.57} &
  \multicolumn{1}{c|}{2025 FEB 04} &
  \multicolumn{1}{c|}{2033 OCT 15} &
  \multicolumn{1}{c|}{\textbf{7.23}} &
  \multicolumn{1}{c|}{13.9} &
  \multicolumn{1}{c|}{11.97} &
  \multicolumn{1}{c|}{12.1} &
  \multicolumn{1}{c|}{3.5} &
  \multicolumn{1}{c|}{89.6} \\ \hline
\multicolumn{1}{|c|}{E-V-r-V-UN$_{271}$} &
  \multicolumn{1}{c|}{2.28} &
  \multicolumn{1}{c|}{2025 MAR 03} &
  \multicolumn{1}{c|}{2033 NOV 11} &
  \multicolumn{1}{c|}{\textbf{7.31}} &
  \multicolumn{1}{c|}{15.8} &
  \multicolumn{1}{c|}{12.23} &
  \multicolumn{1}{c|}{12.16} &
  \multicolumn{1}{c|}{3.34} &
  \multicolumn{1}{c|}{83.14} \\ \hline
\multicolumn{1}{|c|}{E-V-r-V-M-DSM-V-UN$_{271}$} &
  \multicolumn{1}{c|}{1.57, DSM = 2.28} &
  \multicolumn{1}{c|}{2025 FEB 08} &
  \multicolumn{1}{c|}{2033 NOV 25} &
  \multicolumn{1}{c|}{\textbf{7.52}} &
  \multicolumn{1}{c|}{14.34} &
  \multicolumn{1}{c|}{11.97} &
  \multicolumn{1}{c|}{12.19} &
  \multicolumn{1}{c|}{3.73} &
  \multicolumn{1}{c|}{95.13} \\ \hline
\multicolumn{1}{|c|}{E-V-r-V-UN$_{271}$} &
  \multicolumn{1}{c|}{1.57} &
  \multicolumn{1}{c|}{2025 MAR 12} &
  \multicolumn{1}{c|}{2033 NOV 03} &
  \multicolumn{1}{c|}{\textbf{7.81}} &
  \multicolumn{1}{c|}{17.38} &
  \multicolumn{1}{c|}{12.38} &
  \multicolumn{1}{c|}{12.16} &
  \multicolumn{1}{c|}{3.64} &
  \multicolumn{1}{c|}{101.32} \\ \hline
\multicolumn{1}{|c|}{E-M-E-UN$_{271}$} &
  \multicolumn{1}{c|}{N/A} &
  \multicolumn{1}{c|}{2026 NOV 26} &
  \multicolumn{1}{c|}{2033 AUG 07} &
  \multicolumn{1}{c|}{\textbf{7.83}} &
  \multicolumn{1}{c|}{30.48} &
  \multicolumn{1}{c|}{12.65} &
  \multicolumn{1}{c|}{11.9} &
  \multicolumn{1}{c|}{2.31} &
  \multicolumn{1}{c|}{102.16} \\ \hline
\multicolumn{1}{|c|}{E-V-r-V-M-DSM-V-UN$_{271}$} &
  \multicolumn{1}{c|}{1.57, DSM = 2.92} &
  \multicolumn{1}{c|}{2025 MAR 18} &
  \multicolumn{1}{c|}{2033 NOV 27} &
  \multicolumn{1}{c|}{\textbf{9.86}} &
  \multicolumn{1}{c|}{35.26} &
  \multicolumn{1}{c|}{12.57} &
  \multicolumn{1}{c|}{12.19} &
  \multicolumn{1}{c|}{3.92} &
  \multicolumn{1}{c|}{103.17} \\ \hline
\multicolumn{1}{|c|}{E-V-r-V-E-DSM-V-UN$_{271}$} &
  \multicolumn{1}{c|}{1.57, DSM = 2.28} &
  \multicolumn{1}{c|}{2025 JAN 02} &
  \multicolumn{1}{c|}{2033 OCT 03} &
  \multicolumn{1}{c|}{\textbf{10.32}} &
  \multicolumn{1}{c|}{8.4} &
  \multicolumn{1}{c|}{12.22} &
  \multicolumn{1}{c|}{12.07} &
  \multicolumn{1}{c|}{7.42} &
  \multicolumn{1}{c|}{96.96} \\ \hline

\end{tabular}
\end{sidewaystable*}
}

\begin{figure}[ht]
\hspace*{1.0cm}
\includegraphics[scale=0.50]{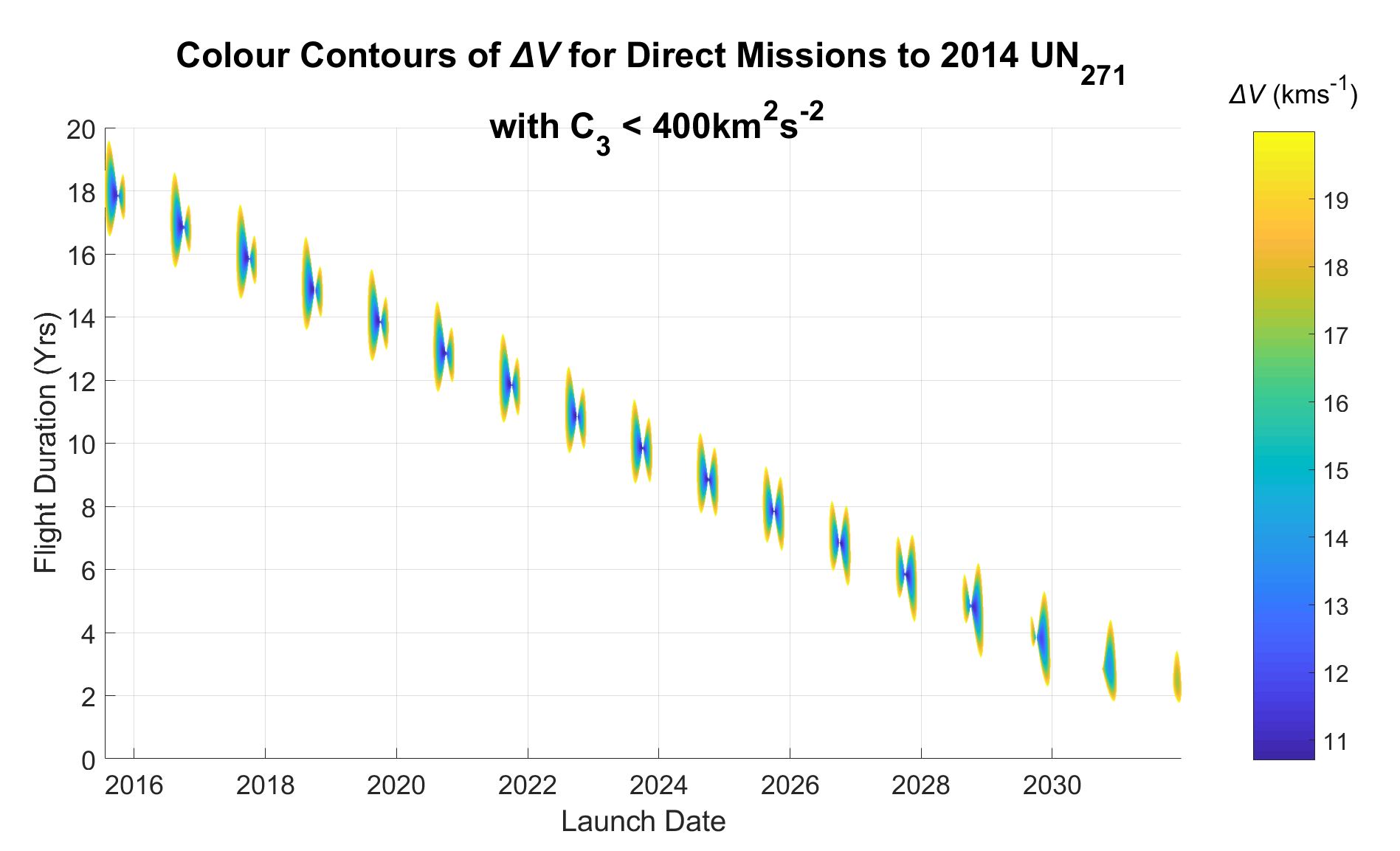}
\caption{White Areas are where Characteristic Energy $C_{3}$ is infeasible}
\label{fig:Direct2}
\end{figure}
In Figure \ref{fig:Direct2} we take Figure \ref{fig:Direct} results and filter out any trajectories for which the characteristic energy required at Earth, $C_{3}$ is too large, specifically $C_{3}$  $>$ 400 $\si{km^{2}.s^{-2}}$. (Note $C_{3}=$ $V_{\infty}^{2}$.) Observe in Figure 2 that there are limited areas of realistic viability for direct flyby missions and they tend to follow a diagonal arrangement on the graph. In fact as we progress from year to year, the feasible flight durations reduce by almost precisely one year, from which we can infer that there is an optimal arrival date at the object UN$_{271}$. Perhaps as one might expect, this date corresponds to when the comet reaches its ascending node with respect to the ecliptic plane, a point at which a spacecraft sent from Earth would \textit{'prefer'} to intercept UN$_{271}$, as it can then use Earth’s own planar velocity optimally to arrive at the target.

Numerical data on this is provided in Table \ref{tab:Direct_T} which gives a summary of the optimal launch scenarios displayed graphically in Figure \ref{fig:Direct2}. Thus for launches between 2022 and 2027 inclusive, the optimal arrival date turns out to be 6 August 2033. For a launch date 2029 and beyond, there is simply insufficient time to intercept UN$_{271}$ at its ascending node and the corresponding $\Delta V$s increase.

\subsubsection{Indirect Transfer} \label{fly_indir}
\paragraph{Using Jupiter} \label{fly_indir_J}

By using Jupiter’s mass, a combined Jupiter GA and Jupiter Oberth manoeuvre (in other words a powered GA) can be attempted as a possible strategy by which overall mission $\Delta V$ can be reduced compared to the direct case (Section \ref{flyby_dir}). In this context, we define here the overall mission $\Delta V$ as the sum of $V_{\infty}$ at Earth and the impulsive change in velocity required at perijove to intercept and flyby UN$_{271}$. 
We may inquire as to whether we can produce a colour contour plot for the mission here, i.e. using a Jupiter encounter, similar to that provided for the direct case in Section \ref{flyby_dir}. The context here is more complicated in that there are two legs which combine to make the overall mission duration, the leg from Earth to Jupiter and that from Jupiter to UN$_{271}$. However if we choose the optimal ratio of times for these legs (in terms of minimizing the previously defined $\Delta V$) then we can construct a plot of the kind we are after. Thus refer to Figure \ref{fig:Jup_1}. Please note in Figure \ref{fig:Jup_1} the following:
\begin{figure}[ht]
\hspace*{-1.0cm}
\includegraphics[scale=0.31]{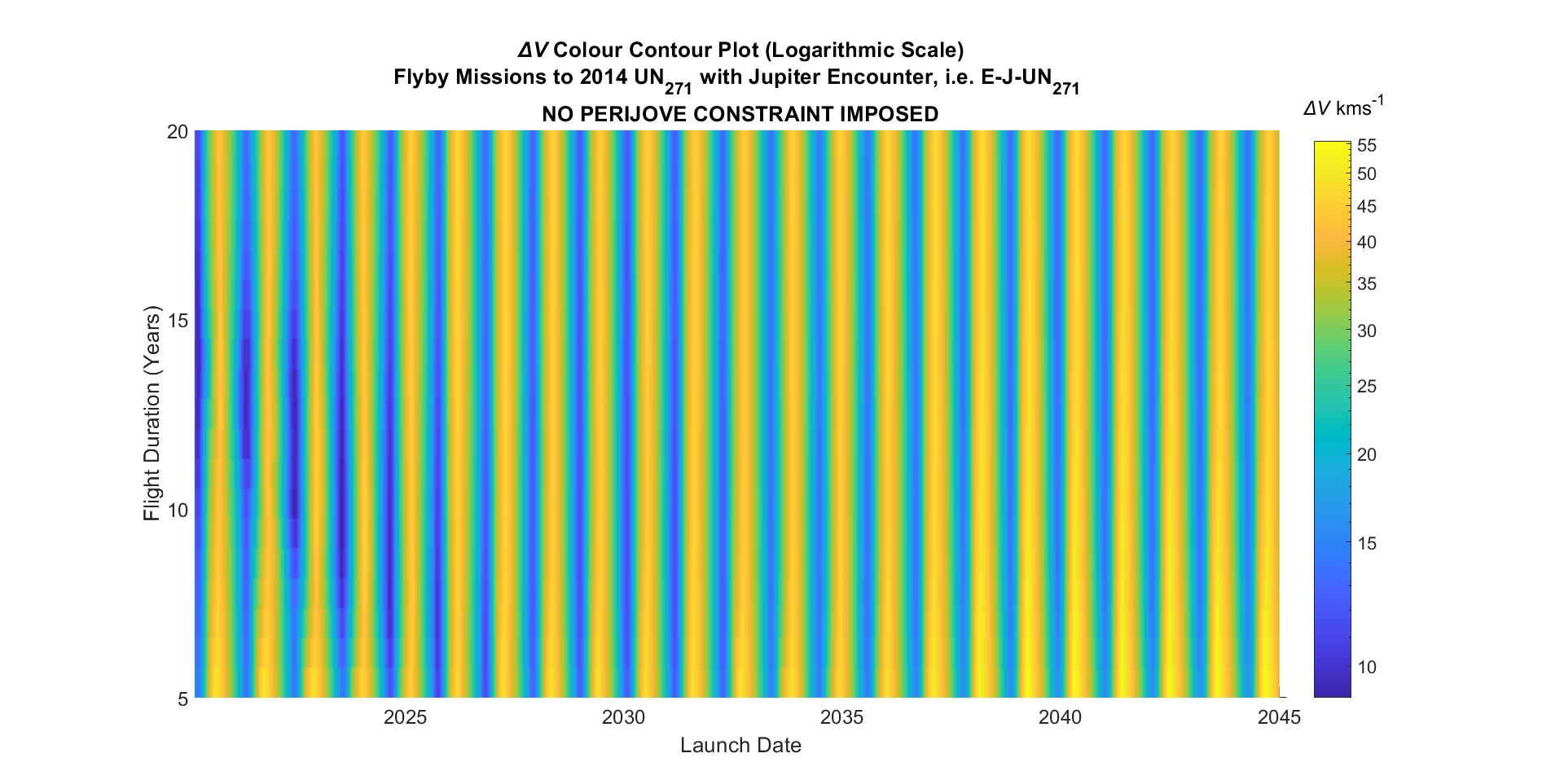}
\caption{Pork Chop plot of missions to flyby UN$_{271}$ employing a Jupiter Flyby}
\label{fig:Jup_1}
\end{figure}
\begin{figure}[ht]
\hspace*{-1.0cm}
\includegraphics[scale=0.28]{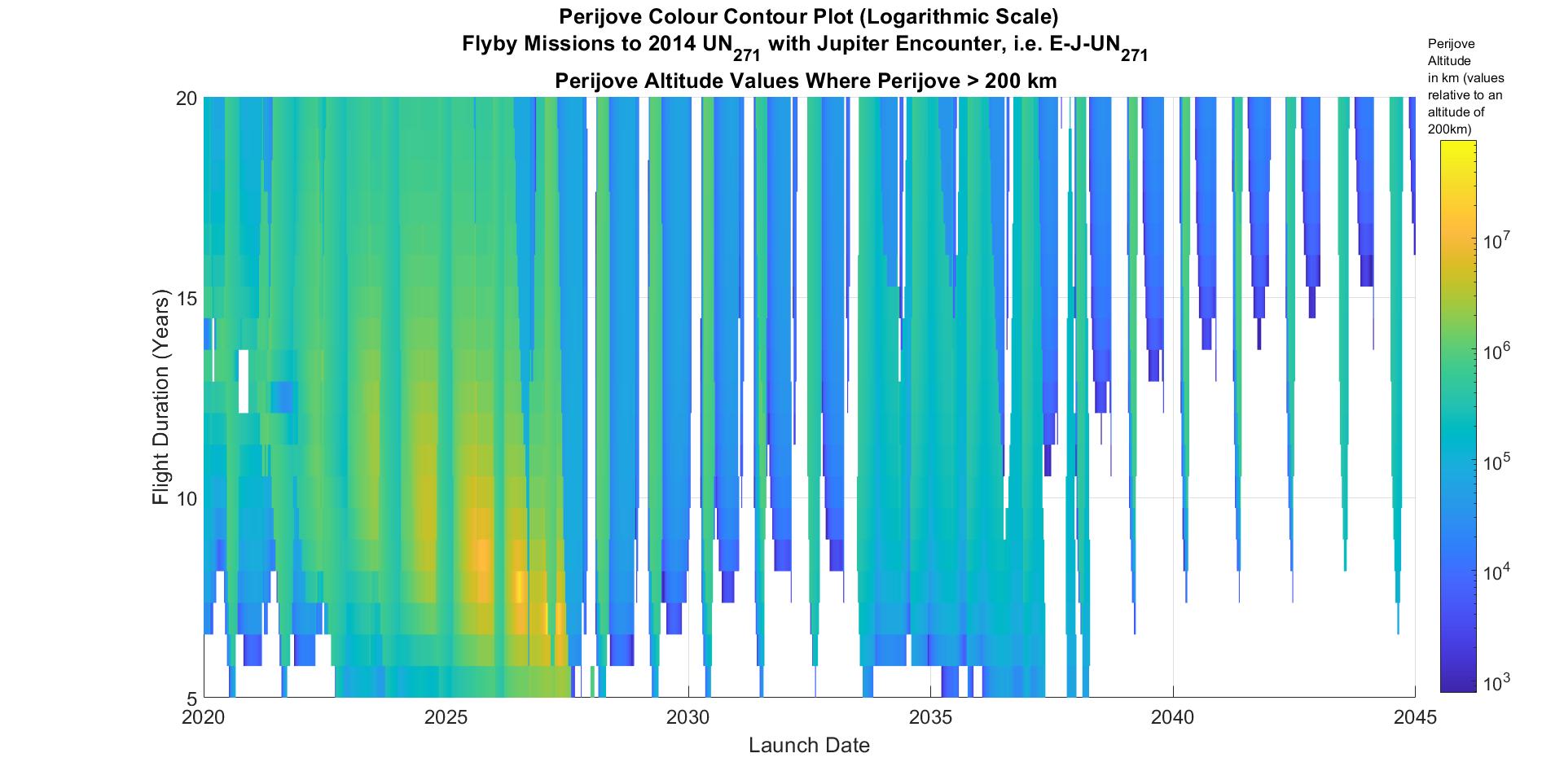}
\caption{Pork Chop of perijove altitudes with respect to 200 $\si{km}$}
\label{fig:Jup_2}
\end{figure}
\begin{figure}[ht]
\hspace*{-1.0cm}
\includegraphics[scale=0.28]{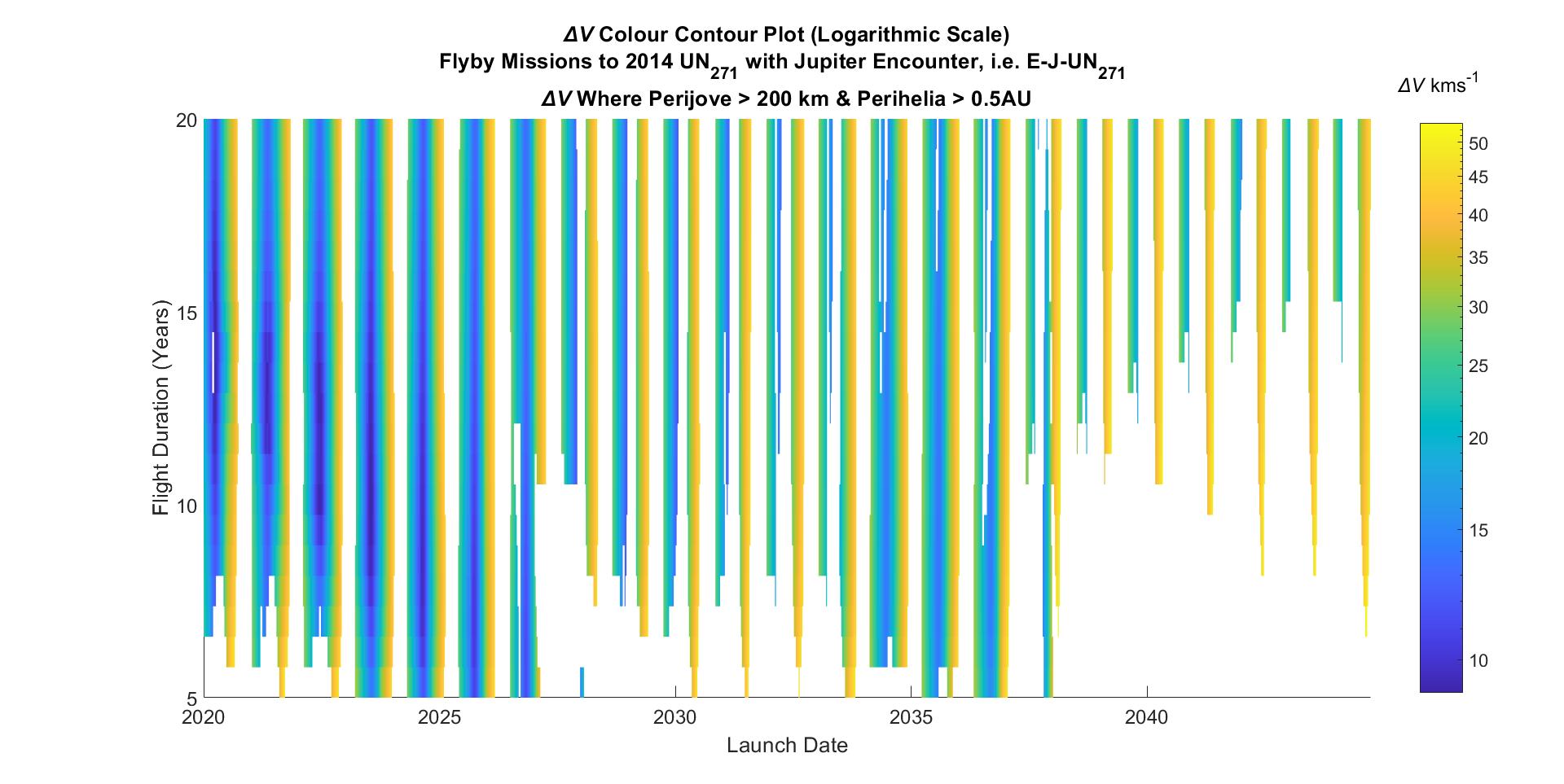}
\caption{Pork Chop plot of feasible $\Delta V$ for missions to UN$_{271}$ with Jupiter Flyby}
\label{fig:Jup_3}
\end{figure}
\begin{figure}[ht]
\hspace*{-0.5cm}
\includegraphics[scale=0.41]{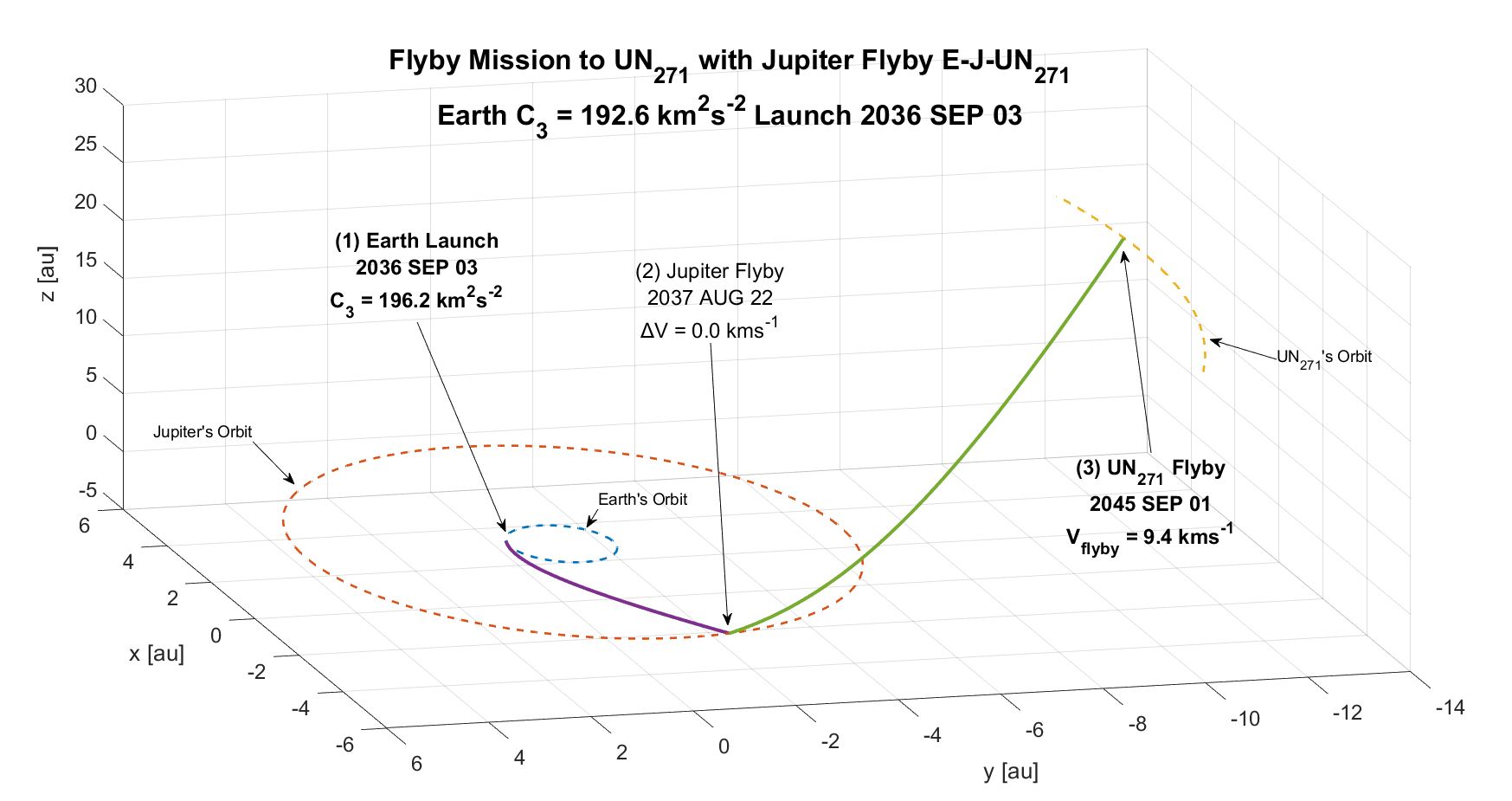}
\caption{Example Mission to UN$_{271}$ with Jupiter Gravitational Assist (GA)}
\label{fig:Jup_4}
\end{figure}
\begin{figure}[ht]
\includegraphics[scale=0.52]{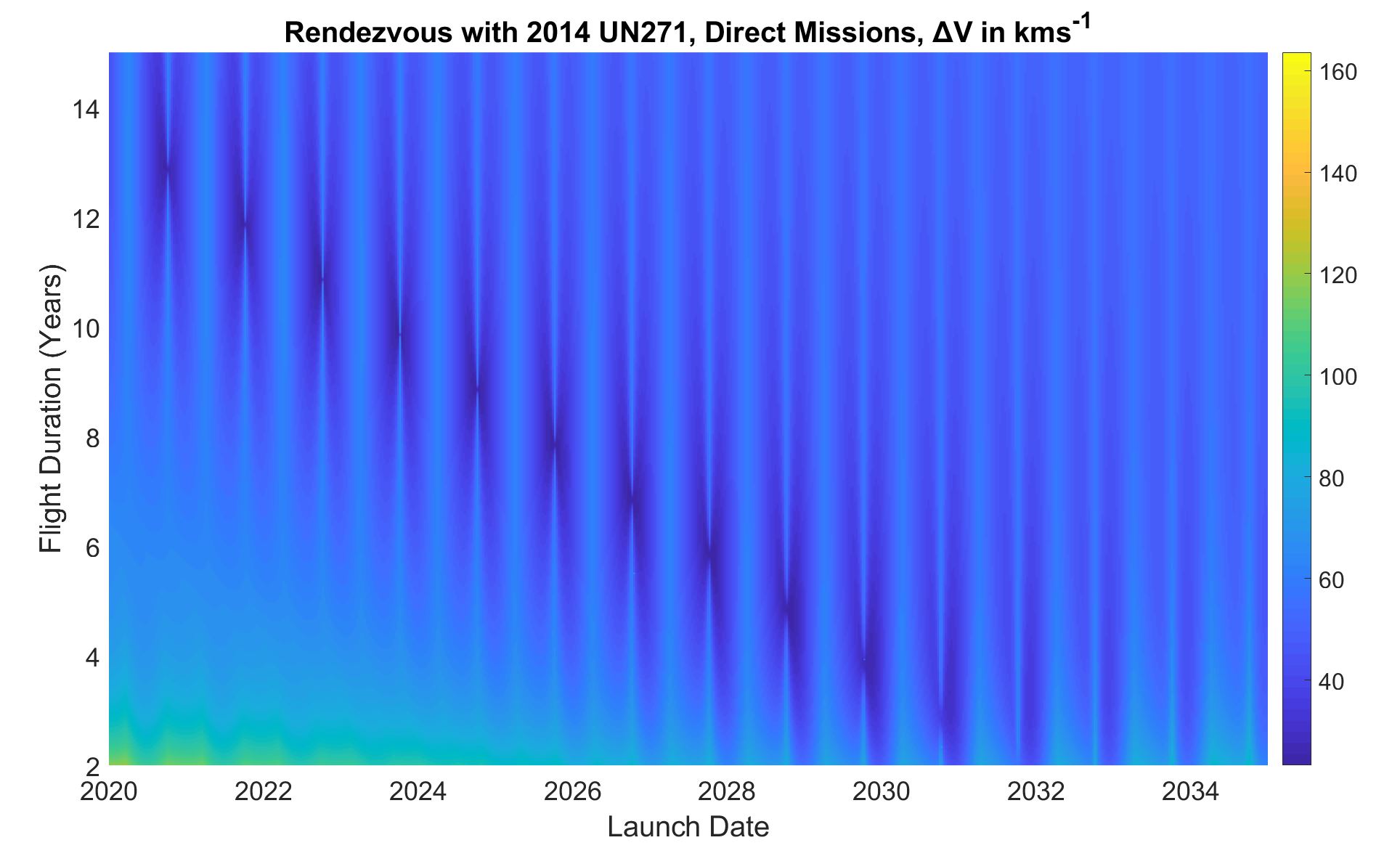}
\caption{Pork Chop Plot for Rendezvous Missions to UN$_{271}$ with colour bar to the right with units in $\si{km.s^{-1}}$.}
\label{fig:Direct_R}
\end{figure}

\begin{figure}[ht]
\hspace*{-1.0cm}
\includegraphics[scale=0.31]{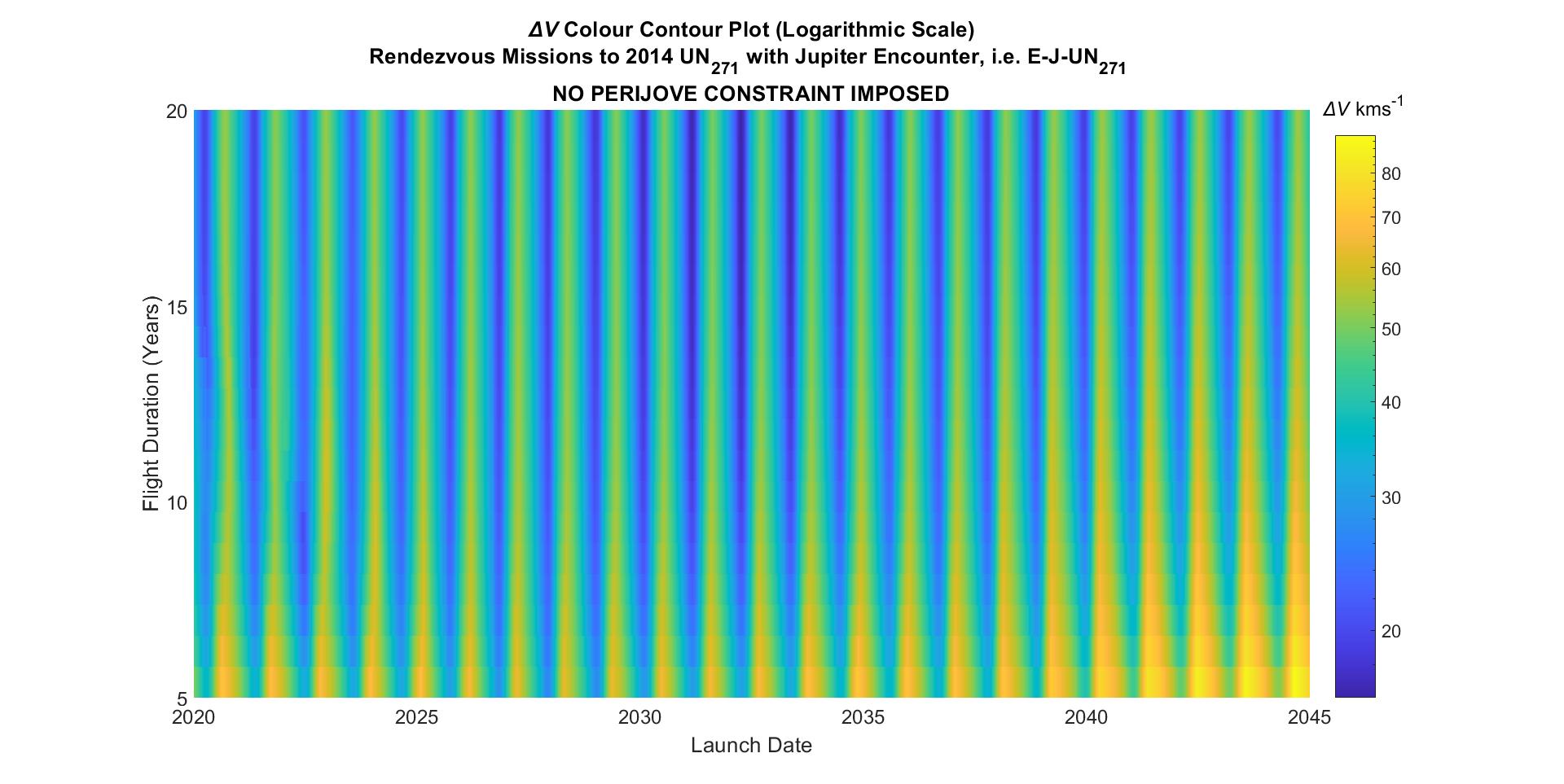}
\caption{Pork Chop plot of missions to Rendezvous with UN$_{271}$ employing a Jupiter Flyby}
\label{fig:Jup_1_R}
\end{figure}
\begin{figure}[ht]
\hspace*{-1.0cm}
\includegraphics[scale=0.28]{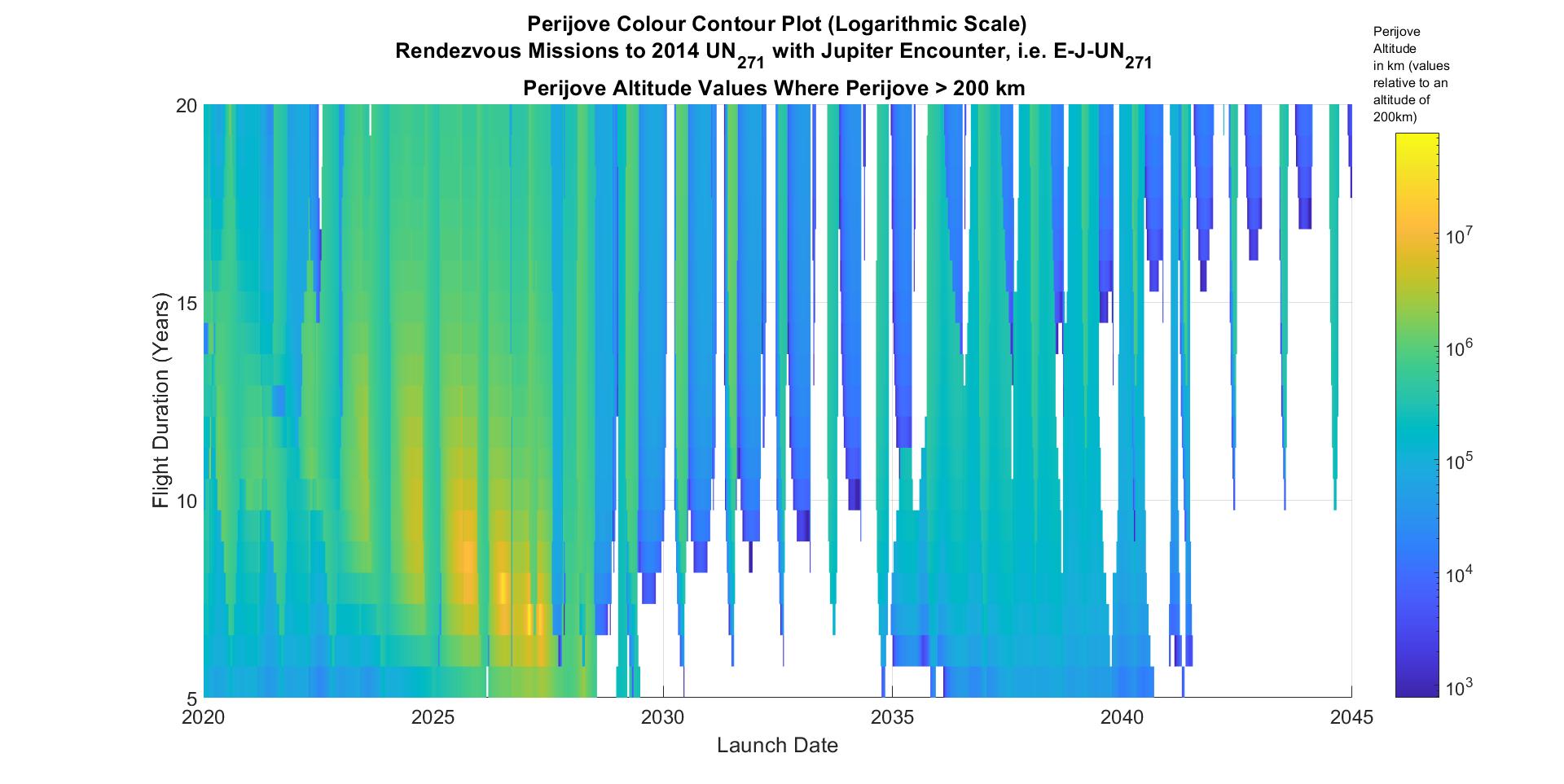}
\caption{Rendezvous at UN271 Pork Chop of perijove altitudes with respect to 200km}
\label{fig:Jup_2_R}
\end{figure}
\begin{figure}[ht]
\hspace*{-1.0cm}
\includegraphics[scale=0.28]{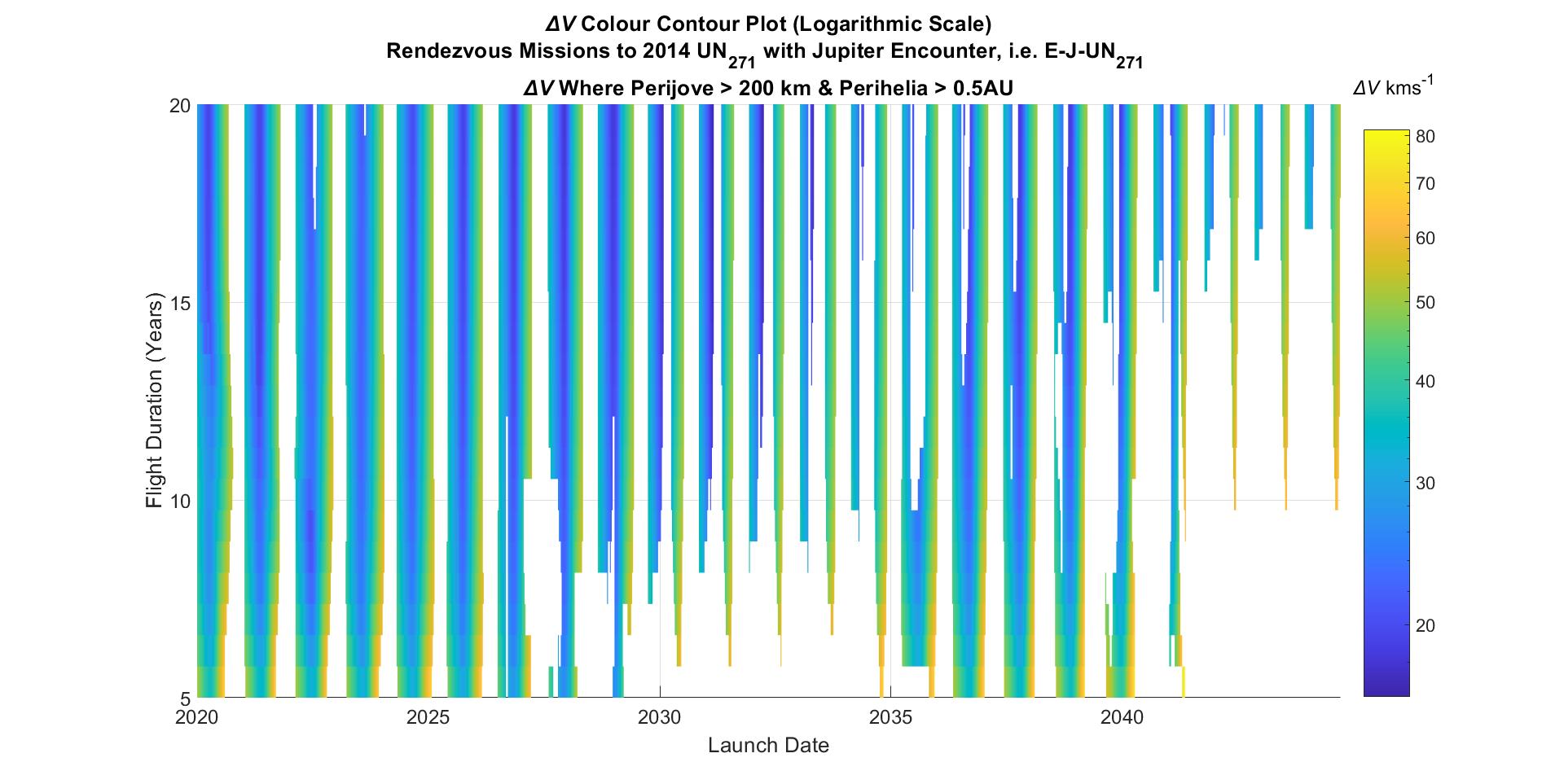}
\caption{Rendezvous Missions. Pork Chop plot of feasible $\Delta V$ for missions to UN$_{271}$ with Jupiter Flyby}
\label{fig:Jup_3_R}
\end{figure}
\begin{figure}[ht]
\hspace*{-1.0cm}
\includegraphics[scale=0.44]{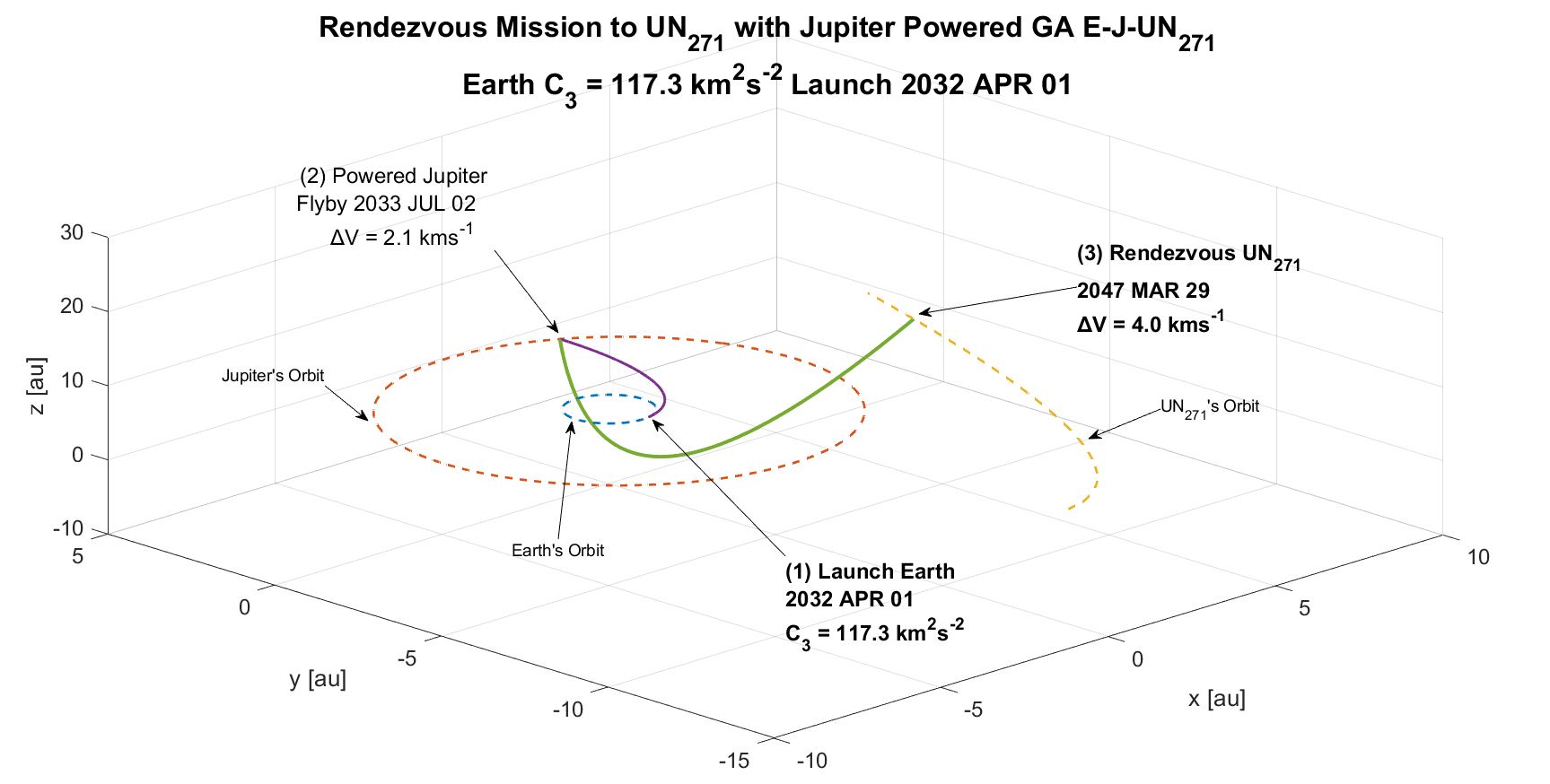}
\caption{Rendezvous with UN$_{271}$, Example Mission with Jupiter GA}
\label{fig:Jup_4_R}
\end{figure}

\begin{enumerate}
    \item There is no constraint on the minimum perijove altitude which the spacecraft can follow, thus this altitude may indeed be negative, obviously making the trajectory impossible to achieve in practice.
    \item A logarithmic scale for $\Delta V$ is provided because of the large range of values this parameter can take.
    \item There is no constraint on the perihelion distance along the interplanetary trajectory, i.e. from the Earth to Jupiter leg or the Jupiter to UN$_{271}$ leg.
    \end{enumerate}

To address first point (1), refer to Figure \ref{fig:Jup_2} which shows the perijove altitudes with respect to a reference altitude of 200 $\si{km}$. Again, due to the wide range of perijove altitudes required (a range of several orders of magnitude), a logarithmic scale is utilized. The white areas or gaps in this plot are where the perijove altitudes $<$ 200$\si{km}$ have been removed.
Figure \ref{fig:Jup_3} takes the white areas in Figure \ref{fig:Jup_2} and removes them from Figure \ref{fig:Jup_1} and further neglects all trajectories for which the perihelion is $<$ 0.5 $\si{au}$. Figure \ref{fig:Jup_3} therefore represents feasible missions in terms of interplanetary trajectories and the orbital mechanics of the Jupiter encounter. Observe that there are two regions of general feasibility from around 2020 to 2027 and then again from 2034 to 2037. In both regions there exist gaps of infeasibility but more importantly, also regions of low $\Delta V$.\\
For an example trajectory involving launch in the year 2022, the minimal $\Delta V$ trajectory involves a launch on 21 JUN 2022, the Earth $C_3$ is 79.2 $\si{km^2.s^{-2}}$  and the arrival speed relative to UN$_{271}$ is 14.5 $\si{km.s^{-1}}$, with a perijove altitude of 660,286 $\si{km}$, and an overall flight duration of around 12 years. This can be compared to the New Horizons spacecraft where $C_3$ = 158.8 $\si{km^2.s^{-2}}$ perijove altitude was 2,300,000 km and the arrival speed was 13.8 $\si{km.s^{-1}}$. In addition the route from Earth to Pluto via Jupiter was accomplished in the span of 9.5 years.\\
Furthermore if we hone in on the second of these intervals, 2034 to 2037, specifically a launch in 2036, with a flight duration constraint of 9 years (approx. that of New Horizons), we find the trajectory shown in Figure \ref{fig:Jup_4}. Note the Jupiter encounter is passive with no applied thrust/$\Delta V$ and the $C_{3}$ at Earth is 192.6 $\si{km^2.s^{-2}}$. This is not only the minimum $C_{3}$ for this flight duration, but as can be observed in Figure \ref{fig:Jup_3}, for launches in the range 2034 to 2037, increasing the flight duration above values of around 9-12 years has no obvious benefit in terms of reducing the Earth $C_3$, which is the main contribution to overall $\Delta V$ (with a passive Jupiter).\\  

\textbf{Note therefore that these results indicate that for a launch in the '30s, a more powerful lift vehicle than used by New Horizons would be required (it used an Atlas V (551) AV-010 + Star 48B 3rd stage)}, in fact even an Atlas V Heavy with Centaur 3rd stage would be inadequate. However, referring to \cite{2021BAAS...53d.451S}, an SLS Block 2 with Centaur upper stage would be capable of doing the job (payload 3.2 $\si{t}$)  or with a CASTOR 30B (payload 1.6 $\si{t}$).\\
Furthermore observe that the alternative of a \textit{direct} mission in the '30s would not be able to achieve intercept in the same time frame as this indirect mission (9-12 years) and with as low a $C_3$ at Earth (refer Table \ref{tab:Direct_T} last row).

\paragraph{Using the Inner Planets} \label{inner_plan}

Various combinations of inner planets were tried to find multiple gravity assist trajectories. Note that there can be no guarantee with OITS as to whether a global optimum is reached, in terms of both (a) finding an optimal solution trajectory in a user-specified sequence of planetary encounters, and (b) finding the globally optimum sequence of encounters. 

In the case of (b), it is the responsibility of the user to specify to OITS the precise combination and sequence of encounters to adopt and it is therefore a question of trying many such combinations, running OITS many times, and concentrating on those which are the best in terms of minimizing $\Delta V$. Table \ref{tab:InDirect_T} provides a summary of the results. Note a preceding 1 year leveraging manoeuvre, with a launch 1 year before the optimal launch dates provided would generally serve to reduce the overall $\Delta V$ for the trajectories and also to reduce the $C_{3}$ needed at Earth to 0.0 $\si{km^{2}.s^{-2}}$.\\

The lowest total $\Delta V$ of 5.54 $\si{km.s^{-1}}$, and at the head of the Table \ref{tab:InDirect_T}, corresponds to the sequence E-V-r-V-E-UN$_{271}$ (with the Venus-Venus segment incorporating a Deep Space Manoeuvre at a 2:1 resonance with Venus, i.e at 1.57 au). The launch is 2028 MAR 10, and arrival at UN$_{271}$ is on 2033 OCT 18. Observe the in-flight $\Delta V$, 2.32 $\si{km.s^{-1}}$ for this mission is the lowest of all missions except for E-M-E-UN$_{271}$ (mission 9 in Table \ref{tab:InDirect_T}) where it is 2.31 $\si{km.s^{-1}}$. For the E-V-r-V-E-UN$_{271}$ mission, the $\Delta V$ is distributed between the encounters and does not exceed 1.0 $\si{km.s^{-1}}$ whereas for the E-M-E-UN$_{271}$ mission, all the 2.31 $\si{km.s^{-1}}$ $\Delta V$ is applied at the Earth return. It is possible therefore that the latter would be the preferred scenario for a flyby mission to UN$_{271}$.The launch would be 2026 NOV 26 with arrival on 2033 AUG 07. \\
For animations of the trajectories E-V-r-V-E-UN$_{271}$ and E-M-E-UN$_{271}$ go to \cite{CiteTrajAnim1} and \cite{CiteTrajAnim2} respectively, where the latter is preceded by a 1 year $V_{\infty}$ Leveraging Manoeuvre.

\subsection{Rendezvous Missions} \label{rendez_res}

Unlike flyby missions, covered in Section \ref{flyby_res}, rendezvous missions involve an application of thrust as the target is approached to match velocities with it.

\subsubsection{Direct} \label{rendez_dir}
A colour contour plot is provided in Figure \ref{fig:Direct_R} which is analogous to Figure \ref{fig:Direct} in Section \ref{flyby_dir}, but in addition to the $V_{\infty}$ needed at Earth, has an extra $\Delta V$ at the target, UN$_{271}$, in order to rendezvous with it. The main conclusion which can be drawn from this, is that the total $\Delta V$ required for this is much larger, and in fact always exceeds 20.0 $\si{km.s^{-1}}$.

\subsubsection{Indirect Transfer}\label{rendez_indir}
\paragraph{Using Jupiter} \label{rendez_indir_J}
Analogous contour plots to those used for the flyby case provided in Section \ref{fly_indir} can be constructed for the rendezvous case and are provided in Figure \ref{fig:Jup_1_R}, Figure \ref{fig:Jup_2_R} and Figure \ref{fig:Jup_3_R}.\\
From Figure \ref{fig:Jup_1_R}, we can observe that the theoretical optimal launch dates for a rendezvous mission with a single Jupiter encounter (without filtering out negative perijove altitudes) are around 2030 to 2034 and with flight durations around 14-15 years. From Figure 10 we can see that when we remove negative perijove altitudes and low perihelia, most of the landscape for the region 2030 to 2034 is removed and is in fact infeasible.\\
However there are also regions of feasibility and to take a case in point, we assume a launch in 2032 and flight duration 15 years from Figure \ref{fig:Jup_3_R} and we get the trajectory shown in Figure \ref{fig:Jup_4_R}.
Observe that the $C_{3}$ for this mission is 117.3 $\si{km^2.s^{-2}}$ is achievable by an SLS Block 2 + Centaur Upper Stage with a payload mass of approximately 10.0 $\si{t}$ \citep{2021BAAS...53d.451S}. A STAR 63F could then deliver the 2.1 $\si{km.s^{-1}}$ at Jupiter (with perijove altitude 3049 $\si{km}$) propelling a payload of 3.9 $\si{t}$ toward UN$_{271}$. Finally for the deceleration to rendezvous, a liquid propellant rocket such as $MMH/N_2O_4$  (\emph{specific impulse, $I_{sp}$ = 341 $\si{s}$}) could be deployed leaving 1.18 $\si{t}$ for the remaining spacecraft to study UN$_{271}$. The rendezvous occurs at a solar distance of 29  $\si{au}$, approximately the same distance at which UN$_{271}$ was originally discovered. 

\section{Discussion}
For flyby missions to UN$_{271}$, there are yearly viable direct missions with launches in the '20s and with $C_3$ $<$ 141 $\si{km^2.s^{-2}}$ (less than required by the New Horizons probe). This can be further reduced by a series of GAs such as E-M-E-UN$_{271}$ where $C_3$ = 30.5 $\si{km^2.s^{-2}}$, though this would necessitate a further minor burn en-route.\\
Looking ahead to flyby missions in the '30s, a direct mission is confounded by the high out-of-plane velocity needed by the spacecraft to intercept UN$_{271}$ which will have passed through its ascending node and will have a significant displacement from the ecliptic plane. However with a Jupiter encounter to induce a significant out-of-ecliptic velocity component, missions become feasible in the mid to late '30s, by which time the SLS Block 2, with appropriate 3rd stage, would be able to deliver a spacecraft to UN$_{271}$ in the span of 9 years (equivalent to the flight-time of New Horizons).\\  
For rendezvous missions in the '20s, any of the flyby missions discussed in Section \ref{flyby_res} can be considered and the $\Delta V$ to match velocities with UN$_{271}$ can be calculated from the arrival velocities with respect to UN$_{271}$. These arrival velocities are generally around $\sim$12-13 $\si{km.s^{-1}}$ which is a hard task for a chemical rocket.\\
For rendezvous missions in the '30s, a powered Jupiter flyby would enable, for example, a launch in 2032 again requiring utilisation of an SLS Block 2 this time with a Centaur third stage, and further burns  at Jupiter (by a STAR 63F for instance) and upon arrival (by a $MMH/N_2O_4$ combination).

\section{Conclusion}
If the case for a flyby mission is considered sufficiently rewarding in terms of the achievable scientific return, then such missions can readily be conducted in the '20s (direct and indirect), as well as the '30s (with a Jupiter encounter). In the case of the latter, a powerful launch vehicle, such as an SLS Block 2, would be needed to generate the required $C_3$ and optimal overall flight durations would be between 9-12 years.

For rendezvous, direct missions in the '20s would be stretched by the high approach velocities of the spacecraft relative to UN$_{271}$. Therefore it would be beneficial, both in terms of mission feasibiltiy and increased mission preparation time, to delay the launch to the early '30s, when in addition a Jupiter powered GA could be exploited to reach the destination.

\bibliographystyle{aasjournal}
\bibliography{library,library_Adam_Hibberd,Hein_ISO_Modified_by_Adam_Hibberd}

\end{document}